\documentclass[prb,twocolumn,aps,longbibliography, superscriptaddress,floatfix]{revtex4-1}

\usepackage{xcolor} 
\usepackage[normalem]{ulem}
\usepackage{graphicx}
\usepackage{hyperref}
\usepackage{amsmath} 
\usepackage{amssymb}
\usepackage{physics}
\usepackage{nicefrac}

\usepackage{booktabs}
\usepackage{natbib}
\usepackage{bbm}

\usepackage{array}
\usepackage{multirow}

\definecolor{indigo}{rgb}{0.29, 0.0, 0.51}
\definecolor{RGBgreen}{RGB}{ 18 173 42}

\newcommand{\muc}{\ensuremath{\mu_c}}
\newcommand{\muTB}{\ensuremath{\mu}}

\newcommand{\Zeeman}{\text{$V_\mathrm{Z}$}}
\newcommand{\Rashba}{\text{$\alpha_\mathrm{R}$}}

\newcommand{\Zeemansquared}{\text{$V^2_\mathrm{Z}$}}
\newcommand{\Rashbasquared}{\text{$\alpha^2_\mathrm{R}$}}

\begin{document}
\title{Transcendental momentum quantization in semiconducting Rashba nanowires and zero energy states in their normal and superconducting phase}
\author{Nico Leumer}
\affiliation{Institute for Theoretical Physics, University of Regensburg, 93 053 Regensburg, Germany}
\affiliation{Université de Strasbourg, CNRS, Institut de Physique et Chimie des Matériaux de Strasbourg, UMR 7504, F-67000 Strasbourg, France}
\affiliation{Donostia International Physics Center (DIPC), Paseo Manuel de Lardizabal 4, E-20018 San Sebastián, Spain}
\affiliation{Department of Theoretical of Physics, Wrocław University of Science and Technology, Wybrzeże Wyspiańskiego 27, 50-370 Wrocław, Poland}
\author{Harald Schmid}
\affiliation
{Institute for Theoretical Physics, University of Regensburg, 93 053 Regensburg, Germany}
\affiliation{Dahlem Center for Complex Quantum Systems, Freie Universität Berlin, 14195 Berlin, Germany}
\author{Milena Grifoni}
\affiliation{Institute for Theoretical Physics, University of Regensburg, 93 053 Regensburg, Germany}
\author{Magdalena Marganska}
\affiliation{Institute for Theoretical Physics, University of Regensburg, 93 053 Regensburg, Germany}
\affiliation{Department of Theoretical of Physics, Wrocław University of Science and Technology, Wybrzeże Wyspiańskiego 27, 50-370 Wrocław, Poland}
\date{\today}
\begin{abstract}
We study finite system properties of the canonical low energy model for a semiconducting nanowire with Rashba spin-orbit coupling. The case of an isolated wire as well as of one proximitized by an s-wave superconductor are considered. Already for the normal wire, the presence of spin-orbit coupling  leads to eigenstates of the finite system composed of more than two momentum eigenstates. The quantization condition  for the wavevectors is not that of a quantum box, but given instead by a transcendental equation linking the involved wavevectors. For the wire with superconducting pairing, the presence of  electron and hole channels  complicates the composition of the eigenstates. In this case we derive an approximate quantization condition close to the phase boundary, and a condition for the appearance of exact zero energy states. It can be satisfied both in the topological and in the trivial phase. Both the trivial and topological zero energy states contribute to the linear transport through Andreev reflection and direct transmission processes, with their relative importance depending on the degree of the states' localization at the boundary.
\end{abstract}
\maketitle
\section{Introduction}
Semiconducting nanowires coupled to a superconductor are one of the main platforms currently developed to host Majorana fermions for use in topological quantum computing.\cite{Aasen2016,Flensberg2021} In the last decade they have been vigorously probed experimentally, with the theory descriptions based on simple low energy models or more realistic coarse-grained geometries. The numerical studies have explored for instance the spatial modulation of the proximity effect, the influence of the magnetic field on the original pairing, or the electrostatics of the gated nanowire which results in a non-constant profile of the chemical potential. Numerical models have also shown that a strong coupling between the superconductor and the nanowire renormalizes its physical properties such as the $g$-factor, spin-orbit coupling and the proximity pairing.\cite{Chevallier2013, Antipov2018, Mikkelsen2018, Reeg2018,Escribano:bjn2018,Woods2018,Winkler2019} Recently, InAs/Al heterostructures realized the topological gap protocol, demonstrating a low false-positive rate for non-topological Andreev states mimicking Majorana zero modes.\cite{Pikulin2021, Morteza2025} Further, measurements of the partial fermion parity in double-nanowire devices mark significant progress toward the realization of Majorana-based qubits.\cite{aghaee2025distinct, microsoft2025interferometric}

On the analytical side, the effective one-dimensional (1D) continuum model of the proximitized nanowire continues to attract attention: it has recently been  used to calculate the renormalized $g$-factor of a quantum dot coupled to the Rashba nanowire,\cite{Dmytruk2018} studied with an antiferromagnetic substrate modifying the boundaries of the topological phase,\cite{Kobialka2021}, analyzed in the context of mapping it to the Kitaev chain and classical spin chains\cite{Pan2022} or of generating the superconducting diode effect\cite{Legg2022}. It has recently been used to study the behavior of Majorana states under dissipation\cite{Kumar2024} and under driving in a Floquet setup\cite{Mondal2023,Roy2024}.\\
 Despite all the attention, exact analytical solutions for the energy levels and eigenstates of a {\em finite size} Rashba nanowire are not known, even for the simplest form of a minimal model with constant parameters. In the case of the Kitaev chain, (exact) analytical solutions\cite{Leumer2020} constituted an important benchmark both for the further numerical\cite{Leumer2020Transport} and experimental studies\cite{Ten2025}. We aim with this work to provide analogous solutions for the proximitized Rashba nanowire.\\
 We begin by searching for the conditions at which exact zero energy states can be found in the finite proximitized Rashba nanowire. Like in the Kitaev chain,\cite{Zvyagin2015,Leumer2020} exact zero energy eigenstates occur only along some lines in the parameter space of the chemical potential $\mu$ and Zeeman energy $\Zeeman$.\cite{DasSarma2012,Rainis2013}. 
In addition, in the numerics we find fully extended exact zero energy states also in the topologically trivial phase. The finding that already the most fundamental form of the nanowire model can support trivial zero energy states is yet another fact to be taken into account when trying to distinguish between topological and trivial zero bias peaks in transport experiments.\cite{Prada2020} The zero energy solutions coincide with the boundaries between different fermion parity regions earlier found numerically.\cite{BenShach2015} 
Our analytical results are quantitatively improving on the earlier ones,\cite{DasSarma2012} but for full agreement with the numerics we would need the exact momentum quantization in the proximitized nanowire. Various approximation schemes to obtain the momentum quantization are discussed in this paper.\\
The usual approach to finding the low energy states of a finite superconducting nanowire is to combine the two subgap states from the left and right end of two semi-infinite wires.\cite{Klinovaja2012,Prada2017,Schuray2018} Deriving the quantization rules for the finite proximitized nanowire is indeed a very difficult task -- we show here that even the non-superconducting nanowire does not obey the standard quantum-box quantization. We derive the transcendental quantization condition for the normal wire, and confirm its validity by comparison with numerical calculation of the nanowire energy spectrum. We also find approximate quantization condition for the low energy excitations close to the topological phase boundary at low chemical potential.\\ 
The zero energy eigenstates close to the phase boundary are found to be localized, while deep in the non-trivial topological phase they become extended. We link the spatial profile of the low energy excitations to the nature of the dominant processes contributing to electronic transport through the system. Using Green's functions techniques\cite{RammerSmith1986,Leumer2020Transport} we calculate numerically the linear conductance through the proximitized 1D nanowire with discretized Hamiltonian. Strongly localized boundary states favor Andreev processes, while those with extended profiles enhance the contribution of the direct transmission. 
\\
The work is structured as follows. In Sec.~\ref{section: models} we recall the model of the Rashba nanowire and review its bulk properties. The constraints on the values of the chemical potential and Zeeman term for which zero energy states occur in a proximitized nanowire are derived in Sec.~\ref{section: zero energy constraints} and their approximate solutions are given for the continuum model. In Sec.~\ref{section: finite} we derive the quantization rule for a finite non-superconducting wire with Rashba spin-orbit coupling and the quantization rules for superconducting wire in two special cases, of zero magnetic field and close to the topological phase transition. The decay length of zero energy states is connected in Sec.~\ref{section: transport} to the dominant contributions to the linear conductance. We finish with a short summary of our results in the conclusions.
\section{Models}
\label{section: models}
\subsection{Continuum model}
\begin{figure}[t!]
\begin{center}
\includegraphics[width=0.9\columnwidth]{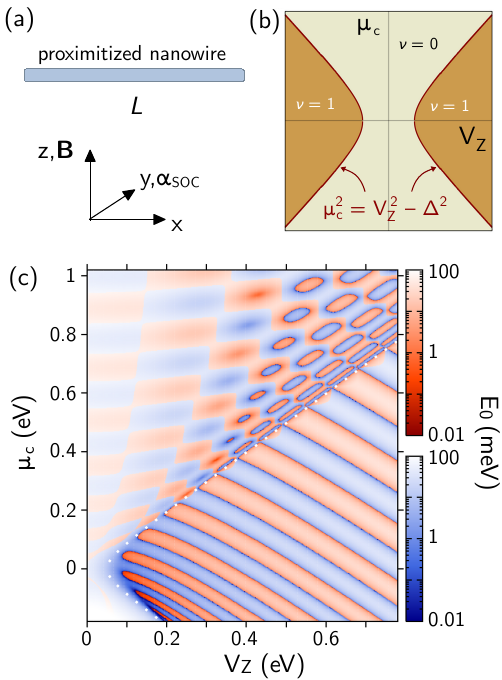} 
\end{center}\vspace{-0.5cm}
\caption{\label{fig: general}
Rashba nanowire. (a) General setup, and (b) Topological phase diagram, where $\nu$ denotes the winding number topological invariant. 
(c) Energy and modified inversion symmetry of the lowest excitation in the numerically solved discretized nanowire with toy model parameters (cf. Tab.~\ref{tab:parameters}) and open boundary conditions. Red (blue) colour scale applies when the lowest excitation is even (odd) under $\tilde{I}$, respectively.  White dotted line marks the boundary of the topological phase for $L\rightarrow\infty$. Note the asymmetry with respect to $\mu_c = 0$, due to the asymmetry of the normal wire's dispersion with respect to $\mu_c$. 
}
\end{figure}

In this section we shortly recall the 1D model used to describe a proximitized Rashba nanowire and its bulk spectrum. The model \cite{Lutchyn2010,Oreg2010}
\begin{align}
\label{equation: wire Hamiltonian}
	\hat{H}_{\mathrm{wire}}= \hat{H}_0 + \hat{H}_\Delta 
\end{align}
is composed of the semiconducting nanowire Hamiltonian
\begin{align}
\label{equation: semi-conducting nanowire Hamiltonian}
	\hat{H}_{0} =\int\limits^{L/2}_{-L/2}\mathrm{d}x&\, \hat{\psi}^\dagger\left(-\frac{\hbar^2\partial_x^2}{2m}-\muc
	-i\, \Rashba\sigma_y\partial_x+\Zeeman\sigma_z\right)\hat{\psi}
\end{align}
and the proximity term $\hat{H}_\Delta$. Here, we introduced the abbreviations $\hat{\psi} =(\hat{\psi}_\uparrow,\hat{\psi}_\downarrow)^\mathrm{T}$, where $\hat{\psi}_\sigma$ removes an electron with effective mass $m$ and spin $\sigma=\,\uparrow,\,\downarrow$ from the nanowire of finite length $L$. The chemical potential is denoted by $\muc$ (where $c$ stands for "continuum") and $\Rashba$ is the strength of the Rashba spin-orbit coupling (SOC), favoring the alignment of spin in the $y$ direction (cf. Fig \ref{fig: general} (a)). 
The heterostructure is exposed to a homogeneous, external magnetic field $\boldsymbol{B}=B\hat{z}$ yielding a Zeeman energy $\Zeeman=g\mu_B B/2$. Due to the proximity effect of the s-wave superconducting substrate, the nanowire becomes itself superconducting, 
\begin{align}\label{equation: superducting pairing term in spin space}
	\hat{H}_\Delta = &\int\limits^{L/2}_{-L/2} \mathrm{d}x\,\big(\Delta\,\hat{\psi}_\uparrow\hat{\psi}_\downarrow+\Delta^*\,\hat{\psi}_\downarrow^\dag\hat{\psi}_\uparrow^\dag\big).
\end{align}
In the following we shall use $\Delta \in \mathbbm{R}$, exploiting the gauge freedom in one superconductor.
In the momentum space, after the conversion to Bogoliubov-de Gennes (BdG) form with the Nambu spinor $ \hat{\Psi}^T(k) = [\hat{\psi}_\uparrow(k),\hat{\psi}_\downarrow(k),\hat{\psi}_\uparrow^\dag(-k),\hat{\psi}_\downarrow^\dag(-k)]^T$, this Hamiltonian can be written as $\hat{H}_{\textnormal{wire}} = \frac{1}{2}\sum_k \hat{\Psi}^\dag(k)\, H_{\textnormal{BdG}} \hat{\Psi}(k)$, where
\begin{align*}
H_{\textnormal{BdG}} = \;&\eta_z\otimes\left[\left(\frac{\hbar^2k^2}{2m} - \mu_c\right)\,\sigma_0 + \alpha_R k \sigma_y + V_Z \sigma_z \right] \\
& + \Delta \eta_y\otimes\sigma_y,
\end{align*}
where $\sigma,\eta$ are Pauli matrices in the spin and particle-hole space, respectively.
This Hamiltonian has neither the conventional inversion nor the time-reversal symmetry, as evidenced by the presence of SOC and of the magnetic field. It does however have a modified inversion symmetry, $\tilde{I} = \eta_z\otimes\sigma_z I$, where $I$ is the spatial inversion symmetry $x\rightarrow -x$ (or $k\rightarrow -k$).
$\hat{H}_{\textnormal{BdG}}$ also possesses local symmetries: pseudo-time reversal $\mathcal{T} = \eta_z\otimes\sigma_z\mathcal{K}$, particle-hole symmetry $\mathcal{P} = i\eta_y\otimes\sigma_z\mathcal{K}$ and the chiral symmetry $C = \mathcal{T}\mathcal{P} = \eta_x\otimes\sigma_0$. The operator $\mathcal{K}$ denotes the complex conjugation. The presence of the chiral symmetry allows us to use the winding number topological invariant, $\nu = \int_{\textnormal{BZ}} dk\,\partial_k \det(h(k))$, where $h(k)$ is the upper off-diagonal block of $H_\textnormal{BdG}$ expressed in the eigenbasis of the chiral symmetry operator $C$.\\

Diagonalising $H_{\textnormal{BdG}}$ yields the excitation spectrum $E(k)$ parametrized by the momentum $k$. Explicitly, one finds that
\begin{align}\label{equation: four bands of the entire hamiltonian}
	E^2(k) &= \left(\frac{\hbar^2k^2}{2m}-\muc\right)^2+\Rashbasquared k^2 + \Zeemansquared +\Delta^2\notag\\
	&\quad \pm 2 \sqrt{\left(\frac{\hbar^2k^2}{2m}-\muc\right)^2 \left(\Zeemansquared +\Rashbasquared k^2\right)+ \Zeemansquared \Delta^2 }
\end{align}
has a gap closing at $k=0$, provided $\Zeemansquared=\muc^2+\Delta^2$ is set. This gap closing condition signals a topological phase transition (TPT). In the case of $\Delta\neq 0$, $\Rashba\neq 0$, $L\rightarrow \infty$ the system is in the topological non-trivial phase for
\begin{align}
	\label{equation: topological phase criterion}
	 \Zeemansquared > \muc^2+\Delta^2,
\end{align}
and in the topological trivial phase otherwise\cite{Oreg2010,Lutchyn2010} as illustrated in Fig.~\ref{fig: general} (b).  

\subsection{Tight-binding model on a lattice}
In order to test the reliability of our analytical calculations, we compare them to the numerical results obtained with the effective lattice Hamiltonian \cite{Stanescu-2011}
\begin{align}\label{equation: tight binding Hamiltonian}
	\hat{H}_{\mathrm{TB}} &=-\muTB \sum\limits_{j=1}^N\sum\limits_\sigma\,\hat{c}_{j\,\sigma}^\dagger \hat{c}_{j\,\sigma}+\Zeeman\sum\limits_{j=1}^N\sum\limits_\sigma\,s_\sigma \hat{c}_{j\,\sigma}^\dagger \hat{c}_{j\,\sigma}
	\notag\\
	 &\quad + t\,\sum\limits_{j=1}^{N-1}\sum\limits_\sigma \left(\hat{c}_{j\,\sigma}^\dagger \hat{c}_{j+1\,\sigma}+ \mathrm{h.c.} \right)
	\notag\\ 
	&\quad + \frac{\alpha}{2}\sum\limits_{j=1}^{N-1}\sum_\sigma s_\sigma\left(\hat{c}_{j+1\,\sigma}^\dagger \hat{c}_{j\,\bar{\sigma}} + \mathrm{h.c.} \right) \notag\\ 
	&\quad + \frac{\Delta}{2} \sum\limits_{j=1}^N\sum_\sigma\,s_\sigma\left( \hat{c}_{j\,\sigma} \hat{c}_{j\,\bar{\sigma}} + \mathrm{h.c.}\right),
\end{align}
where $\muTB$ is the chemical potential, $t$ the nearest neighbour hopping amplitude, $\alpha$ the nearest-neighbor Rashba term, $\Delta$ the $s$-wave superconducting pairing constant, $\Zeeman$ is the Zeeman energy and the length of the wire is $L=Nd$. For given $\sigma$, $\bar{\sigma}$ denotes the opposite spin and $s_{\uparrow(\downarrow)} = +\,(-)$. The resulting bulk Hamiltonian in the $k$-space is
\begin{align}\label{equation: tight binding in k}
\hat{H}_{\mathrm{TB}} = \sum_{k,\sigma} \left\{ (2t \cos(kd) - \muTB + s_\sigma\Zeeman)\, \hat{c}_{k\sigma}^\dag \hat{c}_{k\sigma} \right. \notag\\
 - \left. i\,\alpha s_\sigma\sin(kd)\, \hat{c}_{k\sigma}^\dag \hat{c}_{k\bar{\sigma}} + \frac{\Delta}{2}s_\sigma( \hat{c}_{k\sigma} \hat{c}_{\bar{k}\bar{\sigma}}  + \mathrm{h.c.} )\right\}
\end{align}
and its dispersion relation is
\begin{align}
\label{equation: tight-binding dispersion}
E_{\mathrm{TB}}^2(k) &= (2t\,\cos kd-\muTB )^2 + \alpha^2\sin^2kd + \Zeemansquared + \Delta^2 \nonumber \\
 & \pm2\sqrt{(2t\,\cos kd-\muTB )^2(\alpha^2\sin^2kd + \Zeemansquared) + (\Zeeman\Delta)^2}.
\end{align}
The expansion of the tight-binding Hamiltonian in small $k$ yields the continuum model, with the parameters $t = -\hbar^2/(2md^2)$, $\muTB = \muc + 2t$ and $\alpha=\Rashba/d$. Depending on the ratio of the energy associated with spin-orbit interaction, $E_{so} = m\Rashbasquared/\hbar^2$ to the superconducting gap, the wire can be in the weak ($E_{so}<\Delta$) or strong ($E_{so}>\Delta$) SOC regime. We are going to use only the expansion in small $\Delta$, not in $\Rashba/\Delta$ or its inverse, so we mention this distinction only for completeness. In further numerical calculations we will use two different parameter sets, which we name {\em InAs} and {\em toy} wire models, both in continuum and discrete versions. The parameters are listed in Table~\ref{tab:parameters}.
The InAs wire model is meant to simulate a more experimentally realistic nanowire, based on the InAs parameters~\cite{Stanescu-2011}. The length of the wire is then taken to be $L=1600 \,d \simeq 850$~nm, comparable to experimental lengths.\cite{Albrecht-2016,Deng-2016} The decay length of the subgap states, calculated as $\xi := 1/q$ with $q$ given by Eq.~\eqref{eq: q approx}, is in the range of 0-2$L$ for our explored ranges of $(\Zeeman,\muTB)$. \\
For comparison we study also {\em toy} wires, with $L=100\,d \simeq 53$~nm. In order to be able to resolve the subgap features, we need larger $\Delta$; keeping the same ratio $\Rashba/\Delta$, we rescale the spin-orbit strength and superconducting pairing by a factor of 50. The decay length of the subgap states is smaller than $L$. As can be seen, the InAs wire is in the weak SOC, while the toy wire is in the strong SOC regime, but we reiterate that this distinction is not important in our calculations.\\
 
\begin{figure}[t]
\begin{center}
\includegraphics[width=\columnwidth]{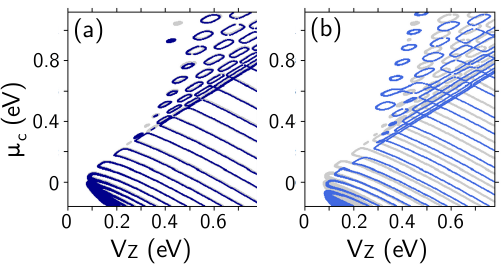} 
\end{center}\vspace{-0.5cm}
\caption{\label{fig: disorder}
Robustness of the zero energy states to disorder in a toy wire. On each site a random potential is added, $\varepsilon_i \in [-W,W]$. The points correspond to $(\Zeeman,\muc)$ parameters where the lowest excitation has energy lower than $2\%$ of $\Delta$. In grey are shown the results from the clean wire, in color those for $W=2\Delta$ (a) or $W=4\Delta$ (b).  While the locations of zero energy states change, they remain stable both in the topological and in the trivial phase. Disorder even widens the region with zero energy states in the trivial phase.
}
\end{figure}

\begin{table*}[htbp]
{\renewcommand{\arraystretch}{1.5}
\begin{tabular*}{\linewidth}{@{\extracolsep{\fill}} l c c c c c c c c c c }
\hline
       &           &             &                & \multicolumn{2}{c}{continuum} & \multicolumn{2}{c}{lattice} & & & \\
 model & $a$ [\AA] & $g\; [\mu_B]$ & $\Delta$~[meV] & $m/m_e$ & $\Rashba$ [eV$\cdot$\AA] & $t$ [eV] & $\alpha$ [meV] & $E_{so}$~[meV] & $\xi(\mu=0,\Zeeman=\Delta)$ & $\xi(\mu=0,\Zeeman=2\Delta)$\\
    \hline
    InAs~\cite{Stanescu-2011} & 5.3 & 15 & 1 & 0.04 & 0.1 & -3.4 & 18.9 & 0.0525 & 100 \AA & 244 \AA \\
    toy & 5.3 & 15 & 50 & 0.04 & 4.32 & -3.4 & 800 & 94 & 85 \AA & 115 \AA \\[2mm]
    strong SOC~\cite{Mishmash2016} &  &  & 0.13 & 0.05 & 0.5 &  &  & 1.636 & 3846 \AA &  \AA \\
    weak SOC InSb~\cite{Mishmash2016} &  &  & 0.25 & 0.015 & 0.2 &  &  & 0.0785 & 800 \AA &  \AA \\
    \hline
  \end{tabular*}
 }\vspace{-0.3cm}
 \caption{\label{tab:parameters} Parameters for the nanowire models used in numerics. The last two models are cited for comparison only.}
 \end{table*}
The diagonalization of the tight-binding Hamiltonian of a finite wire with open boundary conditions shows that Majorana zero modes (MZM) reside on discrete lines within the topologically non-trivial phase as presented in Fig.~\ref{fig: general}(c) for the toy wire  and reported earlier\cite{DasSarma2012,BenShach2015}. Also the topologically trivial phase hosts zero-energy states, on distorted rings in the parameter space.\cite{BenShach2015,Hansen2016} Interestingly, the zero energy states in both phases are resistant to disorder, which on average only changes their location in the parameter space, but does not gap them out, as illustrated in Fig.~\ref{fig: disorder}. We note that the rings don't appear for the lower part of the trivial phase, below $\muc = 0$, since for these chemical potentials all states in the normal wire are empty. They do however extend to much higher values of $\mu_c$ (not shown), in agreement with existing report in the literature.~\cite{Wieckowski2021} Disorder even widens the range of parameters where they are present. In the next section we analyze further the conditions under which they appear. 

\section{Constraints on the zero energy states}
\label{section: zero energy constraints}
Since the topological phase has been extensively studied, we concentrate in the following on the zero energy {\em rings} in the trivial phase, obtaining the zero energy {\em lines} in the topological phase as a by-product.\\

For both $\Zeemansquared\lessgtr\muc^2+\Delta^2$ at $\Rashba =0$, we have that exact zero energy is restricted to parameters obeying the constraint
\begin{align}\label{equation: zero energy lines for zero spin orbit coupling}
	\Zeemansquared = \left(\varepsilon_\Sigma-\muc\right)^2 + \Delta^2,
\end{align}
where $\varepsilon_\Sigma := \hbar^2 k_\Sigma^2/(2m)$ for the continuum or $\varepsilon_\Sigma = 2t\cos(k_\Sigma d)$ for the discrete Hamiltonian, and $k_\Sigma = n\pi/L_+$ (with $n = 1,\ldots,N$, $L_+ /d= N+1$). For generic $\Rashba$, we succeeded to approximate the zero energy conditions for the finite wire length and open boundary conditions, using the full dispersion relation and an approximation for the wavevectors' quantization condition. Below, we proceed with the continuum model. The analogous calculation for the discrete model can be found in Appendix~\ref{appendix: zero energy, TB}.

\begin{figure}[h!]
\includegraphics[width=\columnwidth]{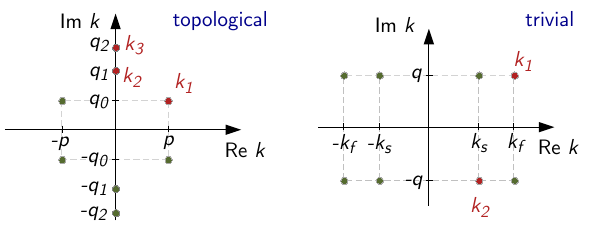}
\caption{\label{fig: momentum roots structure}
Configuration of the solutions of $E^2(k)=0$ in the topologically trivial and non-trivial phase.} 
\end{figure}

The constraint of $E = 0$ in Eq.~\eqref{equation: four bands of the entire hamiltonian} may be written as an 8-th degree polynomial in $k$, in two equivalent forms:
\begin{align}
\label{eq: zero energy polynomial, coefficients}
    0 = P(k) = \sum_{n=0}^8 a_n k^n \equiv \prod_{n=0}^8 (k-k_n), 
\end{align}
where we set $a_8 = 1$. The polynomial $P(k)$ contains only even orders of $k$, therefore its roots will always come in $\pm k_n$ pairs. Moreover, its coefficients are real, therefore the roots will be either real or pairwise complex conjugate. From the existence of the bulk gap we conclude that real roots can exist only on the boundary between topological phases, for $L\rightarrow\infty$. Therefore there are only two available configurations of roots in the complex $k$ plane, shown in Fig.~\ref{fig: momentum roots structure}, with either 4 or 8 roots containing a real part. 
The equality of the two forms of $P(k)$ yields a set of coupled equations relating wavevectors and parameters. For the trivial phase, we set $k_1 = k_f + iq$, $k_2=k_s-iq$ ($k_f,k_s,q\in\mathbb{R}$) granting in the end three independent parameter constraints for zero energy, 
\begin{subequations}
 \label{equation: three equations, continuum}
\begin{align}
 \order{k^6}:\quad & \frac{\hbar^2}{2m}\left( k_f^2+k_s^2-2q^2\right) = 2(\muc + E_{so}) ,\\[2mm]
 \label{eq: cont. model Ok4 equation}
 \order{k^4}:\quad & \left(\frac{\hbar^2}{2m}\right)^2(k_f^2+q^2)(k_s^2+q^2) = \muc^2 + \Delta^2 - \Zeemansquared, \\[2mm]
 \label{eq: cont. model Ok2 equation}
 \order{k^2}:\quad &\left(\frac{\hbar^2}{2m}\right)^3 q^2(k_f^2-k_s^2) = 2\Delta^{2} E_{so}.
\end{align}
\end{subequations}
Notice that $(k_f^2+q^2)(k_s^2+q^2)>0$ implies the restriction to the topologically trivial phase.

Since $m$, $\Rashba$, $\Delta$ are usually fixed for a physical sample, we count in total four variables, namely $k_f$, $k_s$, $q$ and $\Zeeman$ ($\muc$), where the latter depends on $\muc$ ($\Zeeman$). The exact but clearly underdetermined equations Eq. \eqref{equation: three equations, continuum} are completed by the wavevector quantization condition. In this section, as a first step,  we rely on the "particle in a box" quantization; in Sec.~\ref{section: finite} we shall study the quantization condition in more detail.

In the following we use one equation to express $q$ as a function of $\muc$, $\Zeeman$ and the remaining two momenta. After some algebra we are left with two equations, both quadratic in $q^2$. The imaginary part of the wave vector, $q$, controls the decay of the electronic states, $\xi\propto q^{-1}$. From the theory of the Jackiw-Rebbi states~\cite{Shen} we know that $\xi\propto(\Delta E)^{-1}$, where $\Delta E$ is the energy gap in the system. Since our energy gap at $k>0$ is mostly determined by $\Delta$, we can assume that $q$ is small and discard the terms beyond the second order in $q$. (We shall analyze the resulting $q$ in more detail in Sec.~\ref{section: transport}, in the context of electronic transport.) Then, the parameters associated to the zero energy rings obey ($\varepsilon_f:=\hbar^2k_f^2/(2m)$)
\begin{figure*}[hbpt]
\includegraphics[width=0.9\textwidth]{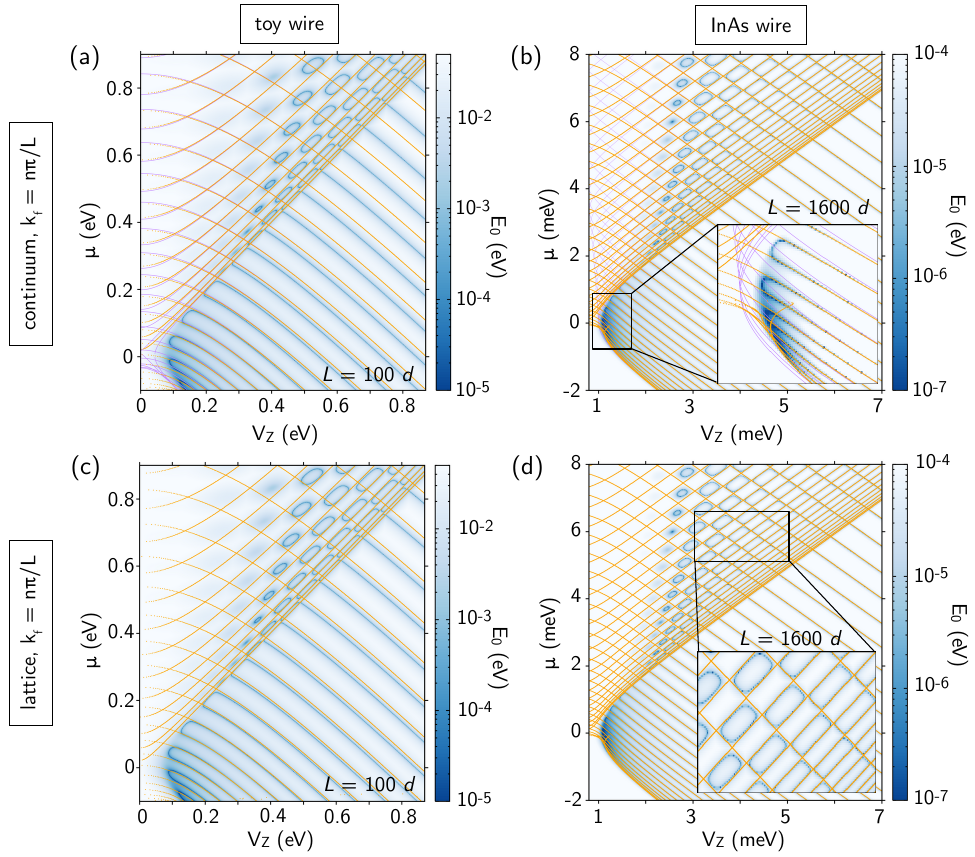} 
\caption{ \label{fig: zero energy lines}
Zero energy lines obtained numerically and with several analytical approaches. The model parameters are given in Table~\ref{tab:parameters}. The orange lines in (a) and (b) are obtained with Eq.~\eqref{equation: zero energy lines continuum}
 for $k_\Sigma = \pi n/L_+$ ($n=1,\ldots,N$). Purple lines are the solutions of an analogous equation $\Zeemansquared = (\varepsilon_f - \muc)^2 - 2E_{so}(\varepsilon_f +\muc) + \Delta^2$ from the supplementary material of Ref.~\onlinecite{DasSarma2012}, plotted for comparison.
 In (c)-(d) we use Eq.~\eqref{equation: lattice vz constraint, first} from the lattice model, together with $k_f = \pi n/L_+$. 
}
\end{figure*} 
\begin{align}
\label{equation: zero energy lines continuum}
\Zeemansquared = (\varepsilon_f - \muc)^2 - 2E_{so}\varepsilon_f + \Delta^2 - \frac{\Delta^2 E_{so}(\varepsilon_f + \muc + E_{so})}{ (\varepsilon_f - \muc - E_{so})^2}. 
\end{align}
In the Figs.~\ref{fig: zero energy lines}(a),(b) we show the energy of the lowest subgap state obtained numerically for the toy (a) and the InAs (b) nanowire. The results of Eq. \eqref{equation: zero energy lines continuum} (assuming that $k_f = n\pi/L_+$ ($n = 1,\ldots,N$)) are shown as orange lines. They capture the first zero energy lines for the toy wire, and all visible lines in the InAs wire. They don't coincide with the zero energy rings in the toy wire, but in the InAs wire the rings are clearly enclosed by the solutions of Eq. \eqref{equation: zero energy lines continuum}. For comparison, we plot with purple lines the zero energy constraint from Ref.~\onlinecite{DasSarma2012}, with accuracy similar to our approximation except for the corner of the topological phase, where our formula captures better the curvature of zero energy lines. In Fig.~\ref{fig: zero energy lines}(c),(d) we show the results of an analogous calculation for the lattice model resulting in  Eq.~\eqref{equation: lattice vz constraint, first} with the same equidistant quantization. 
The accuracy has improved for the toy wire (c)  with respect to the continuum approach, because now the analytical and numerical dispersions match. For the InAs wire the accuracy remains good, with the solutions of Eq.~\eqref{equation: lattice vz constraint, first} only enclosing the zero energy rings, but following closely the topological zero energy lines.
Surprisingly, even though our approach using $k_f,k_s$ and $q$ was meant to capture the zero energy rings in the trivial phase, it does follow closely the zero energy lines in the topological phase, up to the arcs at the topological phase boundary. This unexpected agreement is due to the continuity of the $k_f+iq$ solutions across the upper boundary of the topological phase. The zero energy lines found for equidistant quantization of $k_f$ in the trivial phase are therefore smoothly continued into the topological phase.\\
The zero energy rings in the trivial phase are {\em not} reproduced in either approach, because they both miss the correct, $\Delta$-dependent, momentum quantization. This lack is more visible in the toy wire case, where firstly $\Rashba/t$ is larger than in the InAs wire, and secondly the larger energy window encompasses regions where the band in the discretized model is not parabolic anymore. 
The question of momentum quantization is our focus in the next section, where we obtain the exact quantization rules for the full spectrum of a normal nanowire, for a superconducting nanowire without magnetic field, and an approximate quantization condition for low excitations close to the topological phase boundary.

\section{Finite lattice with open boundary conditions}
\label{section: finite}
The open boundary conditions and the finite length of the nanowire impose, already in the normal conducting case, a non-trivial quantization rule (Eq.  \eqref{equation: quantization rule for zero delta} below) which differs from the typical particle in the box behaviour. Our approach is the usual one of constructing the wave function of a finite system as a linear combination of the bulk Bloch states, imposing open boundary conditions. 
Throughout this section we use the discretized Hamiltonian, so that our analytical results can be directly compared to the numerical ones.

\subsection{The normal conducting wire: $\Delta=0$}
For comparison with the numerical treatment, we consider Hamiltonian Eq.\eqref{equation: tight binding in k} at $\Delta=0$ first. The eigenvectors of $H_{\mathrm{TB}, \Delta = 0}(k)$ are
\begin{align}
v^\pm(k) = N_{k}^\pm\left( \begin{array}{c}
                  i \alpha\sin(kd) \\[2mm] \Zeeman \pm\sqrt{\alpha^2\sin^2(kd) + \Zeemansquared}
                 \end{array}
\right), 
\end{align}
with normalization factor $N_{k}^\pm$. The subscript $\pm$ denotes the sign in Eq. \eqref{equation: tight-binding dispersion} at $\Delta =0$, with these eigenenergies referred to hereinafter as $E_\pm^{\Delta=0}(k)$.

\begin{figure}[htbp]
\begin{center}
\includegraphics[width=0.7\columnwidth]{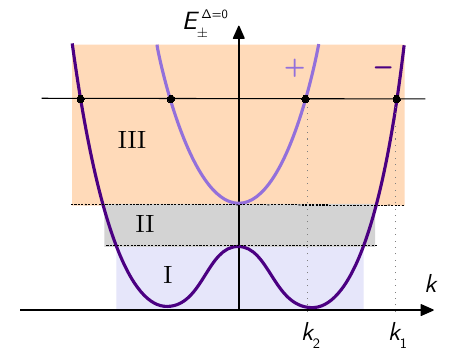} 
\end{center}
\vspace{-0.5cm}
\caption{\label{fig: normal bands}
Sketch of the band structure of a normal wire with magnetic field applied in the $z$ direction. In the region I  four Bloch states are available at each energy, all from the $-$ band. This region shrinks fast with increasing $\Zeeman$. In the Zeeman gap (region II) there are only two Bloch states at each energy, and in the region III again four, one pair from branch $+$ and one from $-$. 
}
\end{figure}
\begin{figure}[htbp]
 \includegraphics[width=\columnwidth]{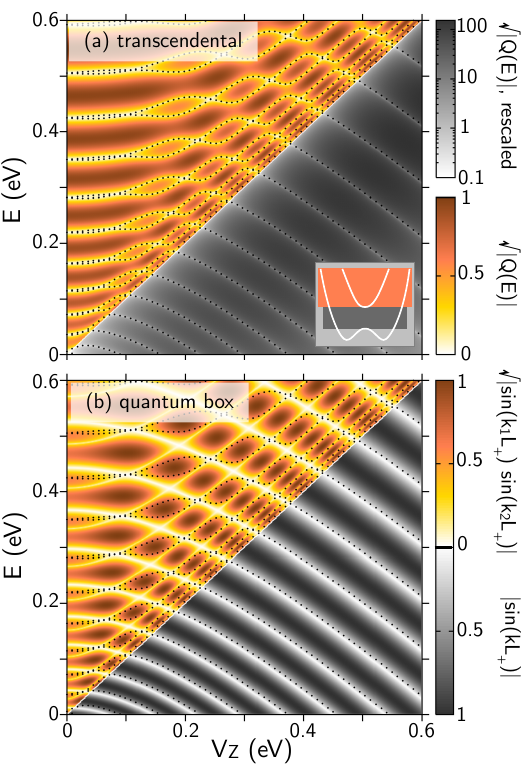}\vspace{-0.2cm}
 \caption{Two approaches to momentum quantization. Black dotted lines mark the energy levels of a discretized {\em toy} nanowire. 
  Background indicates the value of the quantization function, whose minima yield the quantized energy levels of the nanowire: in greyscale for the energy in the Zeeman gap, in color outside of it. (a) Transcendental quantization rule ~\eqref{equation: quantization rule for zero delta}. In the Zeeman gap $Q(E)$ contains exponentially large hyperbolic functions and it has been rescaled by $e^{N(E-V_Z)/|t|}$ for better contrast. Inset shows schematically the relevant energy ranges. (b) Quantum box  quantization function (plotted with square root for better contrast).
  }
 \label{fig: normal quantization}
\end{figure}

The form of $E_\pm^{\Delta=0}(k)$ implies that four wavevectors $\pm k_{1,2}$ correspond to the same energy, cf. Fig~\ref{fig: normal bands}. Thus, we construct $\psi(j)$ as ($x=jd$)
\begin{equation}
\label{equation: eigenstate ansatz}
\psi_{E}(j) = \sum_{n=1,2} A_n v^\pm(k_n)e^{ik_n jd} + B_n v^\pm(-k_n) e^{-ik_n jd},
\end{equation}
with the spinors $v^-(\pm k_1)$ and $v^-(\pm k_2)$ below the Zeeman gap, and  $v^-(\pm k_1)$, $v^+(\pm k_2)$ above it. In regions I and III both $k_i$ are real, within the Zeeman gap one of them is imaginary, corresponding to an evanescent state. \\
The open boundary conditions imply $\psi_{E}(0) = 0 = \psi_{E}(N+1)$. The set of four equations, resulting from two constraints on each of the two components of $\psi_E$, can be written in the form of a matrix equation $M_B \mathbf{A}=0$, where $\mathbf{A}=(A_1,B_1,A_2,B_2)$. Defining $k_\Sigma = (k_1+k_2)/2$, $k_\Delta = (k_1-k_2)/2$, the condition of $\det M_B = 0$ yields the quantization rule
\begin{align}
	\label{equation: quantization rule for zero delta}
	\notag
	Q(E) := & \sin^2(k_\Sigma L_+)\, \left[ 1-\left(\frac{\alpha}{2t}\right)^2\,\cot^2(k_\Sigma d)\right] \\
	- & \sin^2(k_\Delta L_+)\,\left[1-\left(\frac{\alpha}{2t}\right)^2\,\cot^2(k_\Delta d)\right], \notag\\
	\overset{!}{=} & \;0.
\end{align}
This quantization rule does yield the energy spectrum of the finite wire, including the anticrossings caused by the SOC, as we show in Fig.~\ref{fig: normal quantization}(a). The color background corresponds to the value of  $\sqrt{|Q_E|}$, with $k_1,k_2$ found numerically from $E_\pm^{\Delta=0}(k)$ for each energy eigenvalue respectively. Inside the Zeeman gap (greyscale background) two of the wavevectors are imaginary, therefore $Q(E)$ takes exponentially high values between $Q(E)=0$ lines. The numerically found energy levels (dotted black lines) align precisely with the minima of $|Q(E)|$. \\
If we assume from the start the equidistant quantization, we can relate the $k_{1,2}$ to the zeroes of $\sin(k_1 L_+)\sin(k_2 L_+)$. In Fig.~\ref{fig: normal quantization}(b) the color background corresponds to the value of $|\sin(k_1 L_+)\sin(k_2 L_+)|$. Outside of the Zeeman gap its zeroes obviously miss the correct energy levels, but inside the Zeeman gap (greyscale background) there is only one Bloch band at a given energy and the quantization is quantitatively well approximated by that of a quantum box,  $\sin(kL_+)=0$. This justifies the earlier use of $k=\pi n/L_+$ (cf. Fig. \ref{fig: zero energy lines}) to approximate the zero energy lines for a limited $(\mu,V_Z)$ parameter region.\\ 

Even without superconducting pairing, a Rashba nanowire is not just a quantum box, due to the competing influences of spin-orbit coupling and the Zeeman effect on the electronic spin.   
Equation \eqref{equation: quantization rule for zero delta} holds also for $\alpha=0$, $\Zeeman=0$ and, after multiplication by $t^2$, also for $t=0$. It is exact and has to be solved together with the equal energy constraint  $E_\pm^{\Delta=0}(k_1)=E_\pm^{\Delta=0}(k_2)\equiv E^{\Delta=0}$. The constraint allows for alternative representations,
\begin{subequations}
\label{equation: normal wire, equal energy constraints}
\begin{align}
	E^{\Delta = 0} &= -\muTB + \frac{t^2+\frac{\alpha^2}{4}}{t}\,\left[\cos(k_1 d)+ \cos(k_2 d)\right],\\
	E^{\Delta = 0} &= -\muTB  \pm \sqrt{\Zeemansquared +\alpha^2+(4t^2 + \alpha^2)\cos(k_1 d) \cos(k_2 d)},
\end{align}
\end{subequations}
useful for comparison with numerical results. The energy eigenvalues in the limiting cases $\Zeeman \rightarrow 0$, $\alpha \rightarrow 0$ or $t\rightarrow 0$ are presented in appendix \ref{section: Eigenvalues of the tight binding Hamiltonian in limiting cases}.\\
Finally, Eq. \eqref{equation: quantization rule for zero delta} can be obtained by an alternative approach, directly in real space, using the technique of Tetranacci polynomials \cite{Leumer2020, Leumer2020Transport, Leumer2022}. 

We can now check if the position of zero energy lines and rings in the superconducting wire can be found with better precision if we used as $k_f$ or $k_s$ as one of the wavevectors $k_1,k_2$ obeying the quantization condition of the normal wire Eq.~\eqref{equation: quantization rule for zero delta}. (We cannot use both because they would not obey the equal energy constraint with finite $\Delta$). For each $\Zeeman$ we diagonalize numerically the Hamiltonian of the normal wire; using Eq.~\eqref{equation: tight-binding dispersion} at $\Delta =0$ we find the values of $k_f,k_s,q$. Inserting $k_f$ into Eq.~\eqref{equation: lattice vz constraint, second} sets $\muTB$. The resulting zero energy lines and rings obtained are shown in Fig.~\ref{figure: zero energy lines, tight binding}.
\begin{figure}[t]
\begin{center}
\includegraphics[width=\columnwidth]{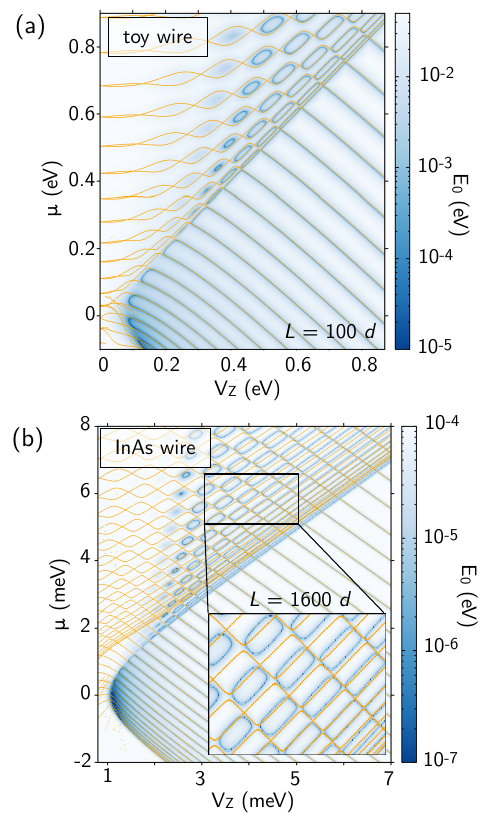}
\end{center}
\vspace{-8mm}
 \caption{\label{figure: zero energy lines, tight binding} 
Zero energy lines and rings with $k_f$ obtained from the wavevectors of a non-superconducting wire (cf. Eq. \eqref{equation: quantization rule for zero delta}). Orange lines mark the solutions of Eq.~\eqref{equation: lattice vz constraint, second}.}
\end{figure}
The use of Eq.~\eqref{equation: quantization rule for zero delta} has strongly improved the agreement between the numerically and analytically found zero energy lines and rings for the toy wire (cf. Figs~\ref{fig: zero energy lines}(c) and \ref{figure: zero energy lines, tight binding}(a)). It does not reproduce the horizontal spacing between rings introduced by $\Delta$ in the presence of the SOC, but it does reproduce their
vertical spacing caused primarily by the SOC. The reason for this good agreement is that the wire is short enough for the size quantization in the normal wire to dominate over the $\Delta$ energy scale and to be the main factor determining the quantization of the superconducting wire as well. In the long InAs wire, with size quantization of the order of $\Delta$, our results follow only the downsloping boundaries of the rings, and the topological zero energy lines. Only using the full quantization condition in the superconducting wire (so far unknown) can yield also the upsloping parts of the rings, and the arcs connecting the zero energy lines in the topological phase (visible in the inset of Fig.~\ref{fig: zero energy lines}(b)).
\subsection{Superconducting wire without magnetic field: $\Delta\neq0$, $\Zeeman=0$}
The s-wave superconducting pairing is, like $\Zeeman$, another factor contributing to the local mixing of the spin degrees of freedom. In the case of $\Zeeman = 0$, spin in the $y$ direction remains a good quantum number and one can separate the Bogoliubov-de Gennes (BdG) Hamiltonian into two distinct $(c_{j\uparrow},c_{j\downarrow}^\dag)$ and $(c_{j\downarrow},c_{j\uparrow}^\dag)$ subspaces, granting rather simple results for the corresponding eigenvalues. The energy eigenvalues of the BdG Hamiltonian at $\Zeeman=0$ read ($n=1,\ldots,N$)

\begin{align}\label{equation: BdG energy spectrum for zero Zeemann}
	E^{\Zeeman=0}_\pm = \pm \sqrt{\left[-\muTB \,+\, 2\tilde{t}\cos\left(\frac{n\pi}{N+1}\right)\right]^2\,+\,\Delta^2},
\end{align}
where $\tilde{t}:=\sqrt{t^2+\alpha^2/4}$ is an effective nearest neighbour hopping constant. Note that the energies in Eq. \eqref{equation: BdG energy spectrum for zero Zeemann} are twice degenerate. 

However, the simplicity of Eq. \eqref{equation: BdG energy spectrum for zero Zeemann} is misleading for the case when all parameters are non zero. 
The dispersion relation of the bulk Hamiltonian, either of the tight binding or of the continuum model, provides eight distinct wavevectors for each $E$, $\pm k_i$ ($i=1,2,3,4$). The ansatz for $\psi(j)$ has to contain these eight wavevectors, in order to satisfy the open boundary conditions. Even accounting for the $\pm k$ symmetry, this leaves us with 4 constraints (particle/hole, left and right end) on 4 variables, twice more than for the normal wire.

The associated general parameter dependent quantization condition is most likely too complicated to be of practical use. In the following we explore only the regime of parameters close to the topological phase boundary, where half-integer momentum quantization has been proposed~\cite{Mishmash2016}, and where an approximate approach could be found.
\subsection{Superconducting wire, close to the topological phase transition}
In this case all parameters $\Delta,\muc,\Zeeman$ are finite, but they obey the condition $\muc^2 \approx \Zeeman^2 + \Delta^2$.
It is natural to expect that close to the TPT the bulk dispersion of the Hamiltonian near the $\Gamma$ point is linear. If this is indeed the case, it can be expressed as a Dirac cone where one parameter is fixed, and the other becomes an effective second momentum component, yielding $E(k,\Zeeman)$ for fixed $\muTB$ or as $E(k,\muTB)$ for fixed $\Zeeman$. From the study of other Dirac systems, such as graphene nanoribbons~\cite{CastroNeto2009} or carbon nanotubes~\cite{Marganska2011} with non-armchair edges, we know that the Hamiltonian may then be split into independent 2x2 blocks with chiral symmetry, and the boundary conditions force the quantization of $k$ to depend on the value of the remaining variable, $\Zeeman$ or $\muTB$. Close to the topological phase boundary, both in the trivial and in the topological phase, we would find $k_n = (n+1/2)\pi/L$ as did the the authors of Ref.~\onlinecite{Mishmash2016}. In Appendix~\ref{appendix: close to TPT} the reader can find a detailed calculation, including the Hamiltonian in a Dirac form close to TPT.\\

\begin{figure}[t]\centering
\includegraphics[width=\columnwidth]{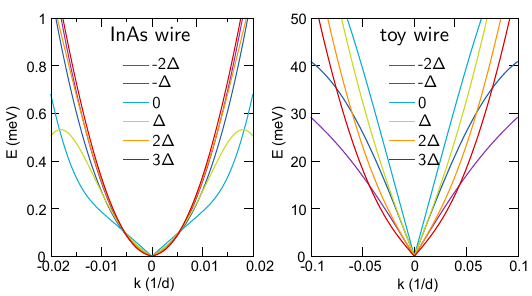}
\caption{\label{fig: dispersion close to TPT}
Bulk dispersion for the long and short nanowire models for several values of $\muTB$ marked by different colors. The linear behavior can be discerned only at $\muTB=0$ in the InAs wire, while for the toy wire it is present in a larger range of chemical potentials, $-2\Delta<\muTB<\Delta$. }
\end{figure}

In order for the dispersion to be linear, the terms quadratic in $k$ have to be neglected -- this is justified only at the values of $\mu$ such that for the relevant values of the momentum the quadratic term is much smaller than the spin-orbit coupling term. Setting this condition somewhat arbitrarily to
\begin{equation*}
    \left\vert\frac{\Rashba \sin k_Fd}{2t(1-\cos k_Fd)}\right\vert \geq 5,
\end{equation*}
we find that in the InAs wire model this corresponds to $\muc \simeq 2\cdot 10^{-5} \,\mathrm{eV}= 0.02\,\Delta$, and for the toy wire to $\muc \simeq 0.05 \,\mathrm{eV}= \Delta$. 
The bulk dispersions for both models at various values of $\mu$ are shown in Fig.~\ref{fig: dispersion close to TPT}, with the linear regime valid indeed only in the close neighbourhood of $\muc\approx0$.
\begin{figure}[htbp]
\begin{center}
\includegraphics[width=\columnwidth]{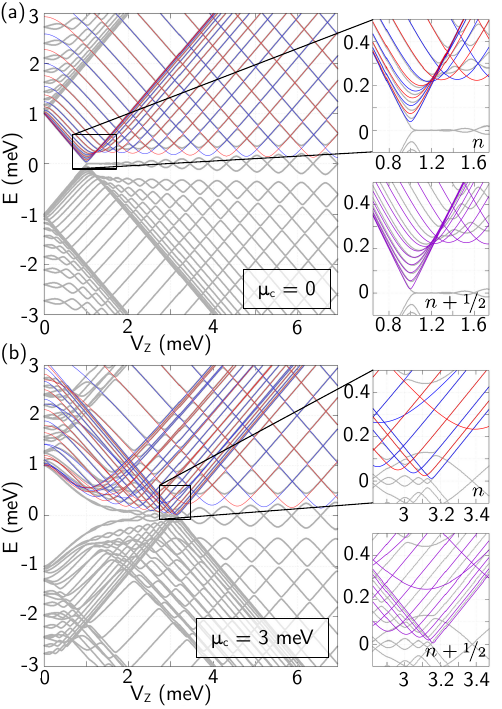}
\end{center}
\vspace{-0.5cm}
\caption{\label{fig: int and halfint} Integer vs half-integer quantization in the InAs model at (a) $\muc=0$ and (b) $\muc=3\Delta$. Numerically calculated spectrum is displayed in grey, colored lines mark the values of $E(k,\Zeeman)$ at quantized values of $k$. Red and blue lines correspond to $k=n\pi/L$ for even and odd integer $n$, purple lines to $k=(n+\nicefrac{1}{2})\pi/L$ values. While in most of the excited states in the topological phase $n\in\mathbb{Z}$ is a good approximation, half-integer $n$ is a better approximation for $\Zeeman$ close to the TPT and $\muc=0$. 
}
\end{figure}
In Fig.~\ref{fig: int and halfint} we show the energy levels of the InAs wire for two values of $\mu_c$ and varying $\Zeeman$. Focusing closer on the range of $\Zeeman$ close to the TPT, we find that at $\mu_c=0$ the half-integer quantization indeed holds, cf. Fig.~\ref{fig: int and halfint}(a) in the zoom panel with $(n+\nicefrac{1}{2})$ quantization. 

While the half-integer quantization reproduces nicely the low energy levels for $\muc=0$, at $\muc=3\Delta$ the integer quantization offers a much better approximation to both the low energy and higher excitation extended states, missing only the anticrossings (cf. the remaining panels of Fig.~\ref{fig: int and halfint}.). The crossings are protected by the modified inversion symmetry $\tilde{I}$ (note that the even or odd $n$ in $k_n$ bears no relation to the even or odd behavior of the state under $\tilde{I}$).
\section{Linear transport}
\label{section: transport}
Transport across a nanowire occurs through a set of channels, provided by its electronic eigenstates. The shape of the wave functions associated with these energy levels determines the type of conduction in the system -- as we will argue, the more an eigenstate resides near the ends of the wire, the stronger is the contribution of Andreev process through this level, at the expense of the direct transmission process. In the following we establish this relation for the zero energy states specifically.
\subsection{Zero energy states - decay length}
Both from Eqs.~\eqref{equation: three equations, continuum} and \eqref{equation: three equations, lattice} we can obtain the inverse decay length, i.e. value of $q$, as a function of $k_f,\Zeeman$ and $\muc$ or $\muTB$, respectively. Here we shall analyze it only in the continuum model. If we again neglect higher than quadratic terms in $q$ in Eqs.~\eqref{equation: three equations, continuum}, we find that
\begin{equation}
\label{eq: q approx}
q^2(\muc,\Zeeman) \simeq \frac{\Delta^2 E_{so}}{\frac{\hbar^2}{m}\left[ (\muc + E_{so})^2 - (\muc^2 + \Delta^2 - \Zeemansquared)\right]},  
\end{equation}
which has no explicit dependence on $k_f$. 

\begin{figure}[htb]
\includegraphics[width=\columnwidth]{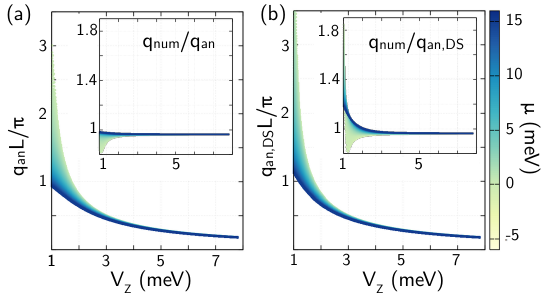}
\vspace{-7mm}
 \caption{\label{fig: inverse decay length} 
  Inverse decay length $q$, in both the topological and trivial phase of the InAs wire, obtained from Eq.~\eqref{eq: q approx} in (a) and, for comparison, from Eq.~(14) in the supplement of \onlinecite{DasSarma2012} in (b), projected against $\Zeeman$. The color denotes the value of $\muc$ for the given data point. The insets show the ratio of the $q$ obtained analytically to that from the numerical solutions of Eq.~\eqref{eq: zero energy polynomial, coefficients}. For large $q$ the analytical approximations worsens as we neglected higher order terms. 
}
\end{figure}

This approximation is very good at high $\Zeeman$ but loses precision close to the tip of the topological phase (low $\muc$ and $\Zeeman$), as we show in Fig.~\ref{fig: inverse decay length}. The values of $\muc,\Zeeman$ where zero energy states occur are found numerically. The value of $q_\text{num}$ is calculated numerically from Eq.~\eqref{eq: zero energy polynomial, coefficients} as the imaginary part of $k_1$ (cf. Fig.~\ref{fig: momentum roots structure}). The purely imaginary wavevectors in the topological phase are not considered. For the same $\muc,\Zeeman$, the value of $q_\text{analytical}$ is found from Eq.~\eqref{eq: q approx}. With decreasing $q$, the decay length increases and the zero energy states become delocalized over the whole 1D wire. As we show in the next section, this determines the dominant processes contributing to zero bias transport through the proximitized Rashba nanowire. 
\subsection{Transport modelling and results}
\label{sec: transport proper}
\begin{figure}[htbp]
\centering
  \includegraphics[width=\columnwidth]{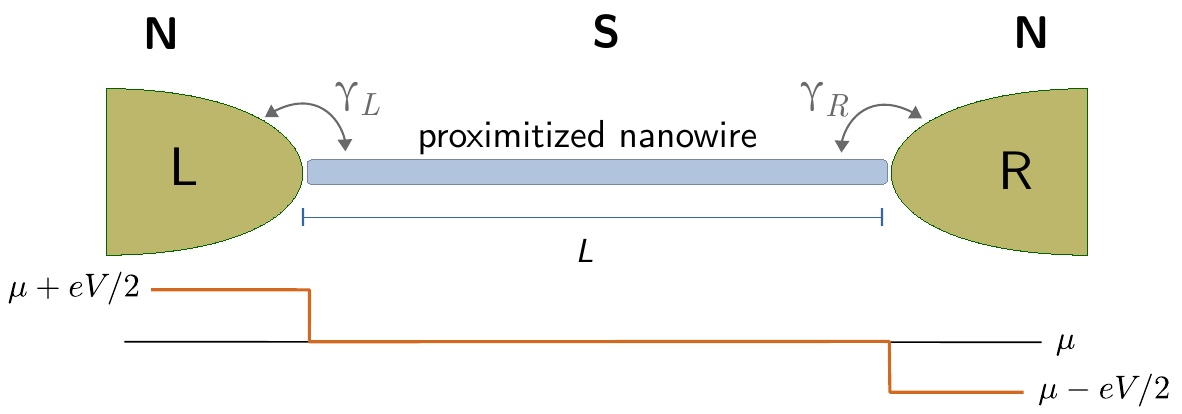}
  \caption{The considered N-S-N transport setup. The nanowire and the proximity-coupled s-wave superconductor form a topological superconductor in the center. Normal conducting leads couple to this central system in the wide-band limit via $\gamma_{L/R}$. Below, the symmetric bias configuration is sketched.}
\label{fig: NSN_structure}
\end{figure}

In this section we introduce our N-S-N transport setup, illustrated in Fig.~\ref{fig: NSN_structure}, and shortly describe the method used to calculate the linear conductance. %
The method is based on the one developed earlier for the Kitaev chain\cite{Leumer2020Transport} in the framework of non-equilibrium Green's functions, extended to a system with additional spin degree of freedom. The leads are assumed to be metallic, and the tunneling Hamiltonian connects the leads to only the first and last site of the nanowire,
\begin{equation}
H_T = \sum_{\sigma} \sum_{k\in L,R} t_1(c_{1\sigma}^\dag d_{k_L\sigma} + c_{N\sigma}^\dag d_{k_R\sigma} + \textnormal{h.c.}). 
\end{equation}

The numerical calculation is performed in direct space for the lattice model. 

With the wide-band limit the coupling to the leads is described by the constant matrices $\Gamma_\alpha$ ($\alpha=L,R$), which contain only four non-zero entries, on the diagonal -- for $\Gamma_L$ they are on the first site of the wire and for $\Gamma_R$ on the last site. The corresponding self-energies are given by $\Sigma^r_\alpha = -i\Gamma_\alpha/2$, with non zero entries $\gamma_L,\gamma_R$, respectively. Transport occurs through Andreev reflection and normal transmission processes. A symmetric bias is chosen such that the current is conserved, i.e. $I_L=-I_R$ \cite{Lambert1993,Lim2012}, without a self-consistent calculation of $\Delta$ \cite{Yeyati1995,Melin2009}. This choice of  the bias drop renders the third possible process (the crossed Andreev reflection) zero, and sets the maximum conductance to $e^2/h$.\cite{Lim2012,UlrichHassler2015,Leumer2020Transport}
\\
In order to shorten the computation time, the Green's function is calculated using a decimation procedure,\cite{Nemec2007} which results in the $8\times 8$ matrix structure 
\begin{equation*}
G^R = \left( \begin{array}{cc} 
                G_{11} & G_{1N} \\ G_{N1} & G_{NN}  
             \end{array}
\right), \quad \text{with }
G_{nm} = \left( \begin{array}{cc}
                 G_{nm}^{pp} & G_{nm}^{ph} \\[2mm] G_{nm}^{hp} & G_{nm}^{hh}
                \end{array}
\right). 
\end{equation*}
The indices $p,h$ denote the particle and hole sectors of the Nambu space. Each of these is again a $2\times 2$ matrix, encoding the spin degree of freedom,
\begin{equation*}
G_{nm}^{xy} = \left( \begin{array}{cc}
                      G_{nm}^{xy,\uparrow\uparrow} & G_{nm}^{xy,\uparrow\downarrow} \\[2mm]
                      G_{nm}^{xy,\downarrow\uparrow} & G_{nm}^{xy,\downarrow\downarrow}
                     \end{array}
\right). 
\end{equation*}  
Following Ref.~\onlinecite{Leumer2020Transport} we find, exploiting the symmetries of the system, that for the linear conductance with symmetric bias drop the transmission is a sum of contributions from Andreev~\eqref{eq:GA} and direct~\eqref{eq:GD} processes,

\begin{align}
G = & \;G_A + G_D, \\[2mm]
G_A = & \;\gamma_L^2 \,\frac{e^2}{2h} \sum_{\sigma=\uparrow,\downarrow} \left(\left|
G_{11}^{ph,\sigma\sigma}\right|^2 + \left|G_{11}^{ph,\sigma,-\sigma}\right|^2 \right), \label{eq:GA}\\[2mm]
G_D = & \gamma_L\gamma_R \frac{e^2}{2h} \left[
\sum_{\sigma=\uparrow,\downarrow}\left|G_{1N}^{pp,\sigma\sigma}\right|^2 + 2\,\text{Re} \left(G_{1N}^{pp,\uparrow\downarrow}\,G_{1N}^{pp,\downarrow\uparrow}\right)
\right]. \label{eq:GD}
\end{align}

\begin{figure}[h!]
\centering
  \includegraphics[width=0.85\columnwidth]{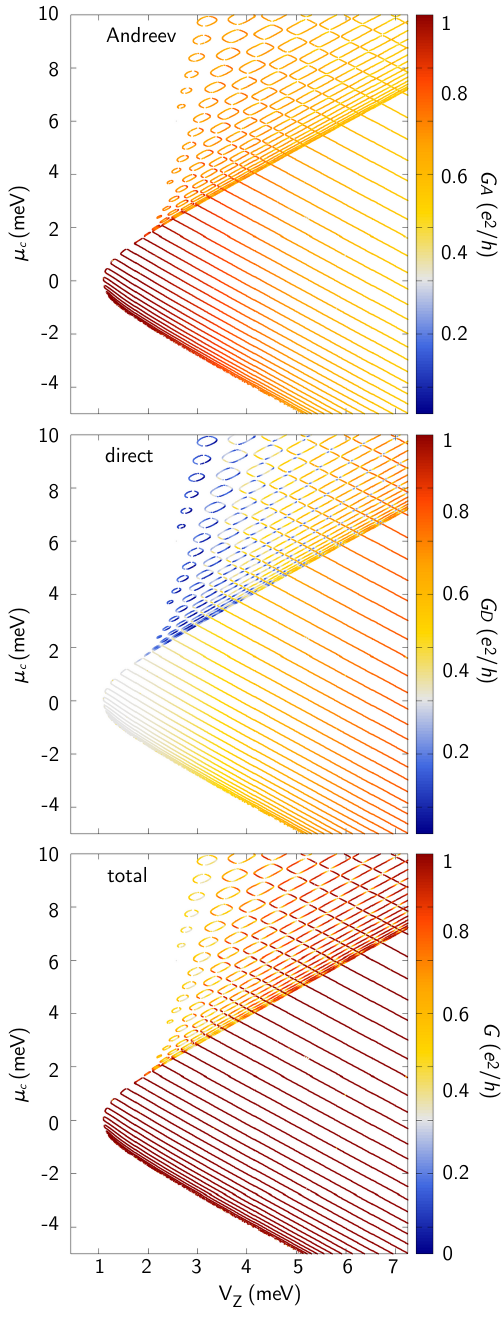}
  \vspace{-5mm}
  \caption{Contributions to linear conductance at the zero energy lines of the InAs wire
  in the topological phase and zero energy rings in the trivial phase. The importance of the Andreev reflection decreases with $\Zeeman$ and that of the direct transmission increases.  
   }
\label{fig: transport contributions}
\end{figure}
The numerical results for these two contributions to the linear conductance in the InAs wire with $L=1600d$ and $\gamma_L = \gamma_R = 0.1$~meV, are shown in Fig.~\ref{fig: transport contributions}. The Andreev term is $\sim e^2/h$ only close to the low $\Zeeman$, low $\muc$ tip of the topological phase, where the localization of the zero energy states is strongest.\cite{Leumer2020Transport} This enhances the amplitude of the wave function on the first site, where the Andreev reflection occurs. The opposite is true for the direct term, which gains in importance as the zero energy states become more extended such that the direct connection to the other end of the chain improves. The total linear conductance stays at $e^2/h$ on the Majorana lines in the topological phase (except for the lines along the lower boundary of the topological phase), and on their extensions into the trivial phase. Elsewhere the conductance is lower than the maximum. 

\begin{figure}[h!]
\centering
  \includegraphics[width=0.9\columnwidth]{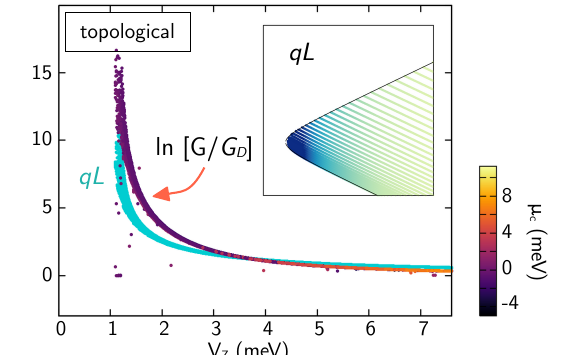}
  \vspace{-2mm}
  \caption{Inverse localization length and $G/G_D$ ratio in the topological regime of the InAs nanowire, projected onto $\Zeeman$ axis for all values of $\mu_c$. The insets indicate the variation of $q$ on the zero energy lines and rings, from dark (high $q$, small localization length $\xi$) to light (low $q$, large $\xi$). The ratio of the wave function amplitudes for zero energy states between the left and right end is $e^{qL}$ and agrees well with $G/G_D$. 
  }
\label{fig: AtoD vs qL}
\end{figure}

\begin{figure}[h!]
\centering
  \includegraphics[width=0.9\columnwidth]{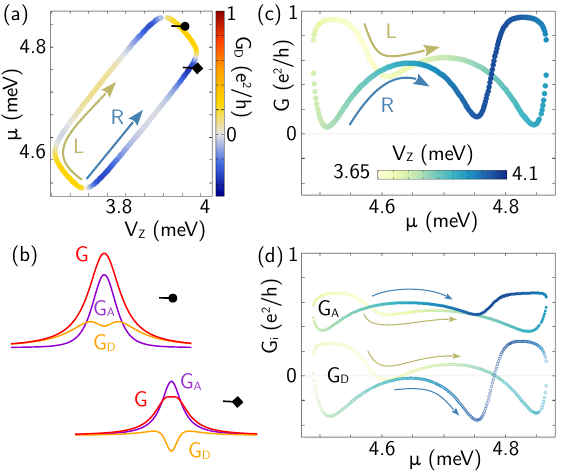}
  \vspace{-2mm}
  \caption{(a) Direct term in the conductance on a ring of trivial zero energy states, on the left (L) and right (R) branch. (b) Cuts along $\Zeeman$ across the zero energy line of $G_A$,$G_D$ and $G$ at positions marked in (a). (c) Evolution of conductance along the L and R branches, with color denoting the value of $\mu$. (d) The same but split between the Andreev and direct contributions, showing clear regions with negative $G_D$. 
  }
\label{fig:ring transmission}
\end{figure}

Both the inverse localization length $q$ and the ratio of total conductance $G=G_A+G_D$ to direct transmission are shown in Fig.~\ref{fig: AtoD vs qL}. In the topological phase, where the oscillation of the the zero energy states' wave function is governed only by one $\Re(k)$ (cf. Fig.~\ref{fig: momentum roots structure}), the wave functions at the left and right end are given (up to a normalization factor) by $\psi(x=0)=1$ and $\psi(x=L)=e^{-qL}$. This ratio governs also the ratio of the Andreev and direct transmission contributions to the conductance.
This estimate doesn't hold for the zero energy rings in the trivial phase, where there are two interfering oscillating components, with $k_f$ and $k_s$ (see Fig.~\ref{fig:ring transmission}). The transmission through the zero energy states on the trivial ring varies from 0 to 1, resulting from an interplay between the strong Andreev contribution and the direct term which can turn negative due to interference of the spin-flipping terms caused by the presence of the spin-orbit coupling. Note that this happens only on the parts of the ring where the value of $k_f$ changes (positive slope in the $(\Zeeman,\mu)$ plane, see also Fig.~\ref{fig: zero lines - normal spectrum}).
\section{Conclusion}
\label{section: VIII conclusion}
Despite extensive interest in semiconducting Rashba nanowires—both as standalone systems and as building blocks for complex heterostructures—exact analytical solutions for the energy spectrum and eigenstates of finite-length nanowires subject to open boundary conditions remained notably absent. In this work, we provide these results, both for the continuous and discrete formulations of the minimal model for semiconducting nanowires.

Although, we present exact results for limiting cases of vanishing spin-orbit coupling $\alpha =0$, vanishing hopping $t=0$ or external magnetic field $\Zeeman = 0$ for the lattice model, one of our main results is the transcendental quantization condition for the wavevectors of the finite wire at $\Delta =0$.

For the superconducting wire $\Delta\neq 0$, we investigate the appearance of exact zero energy states respecting finite size and open boundary conditions. The comparison between two forms of the zero energy condition yields a set of equations, whose approximate solutions agree very well with the results of numerical calculation of the energy spectrum. We exploited that $\Delta$ is typically the smallest energy scale such that wavevectors for $\Delta \neq$ are essentially those at $\Delta =0$. Our solutions reproduce the zero energy lines in the parameter space both in the topological and give a good approximation in the trivial phase. Only the knowledge of the full, $\Delta$-dependent quantization condition would allow us to reproduce also the ring structure of the zero energy features in the trivial phase. We benchmarked our results against approximate literature methods, clarifying their limitations and the advantages of our approach.

Motivated by the role of spatial profiles in quantum transport\cite{Leumer2020Transport}, we analyzed the linear conductance of a finite wire under open boundary conditions. Zero-energy states in both the trivial and nontrivial topological phases generally contribute through Andreev reflection and direct transmission for a symmetrically applied bias which ensures current conservation. In the non-trivial phase and while the total conductance remains quantized, the proportion of Andreev reflection to direct transmission contribution is determined mostly by the degree of localization of the zero energy states near the wire boundary.

Our results advance the understanding of a basic model used to describe one of the best developed experimental platforms for topological superconductivity. Finding zero energy modes in the trivial phase, already in the most fundamental form of the model, is undoubtedly important for the continuing efforts to distinguish between trivial and topological states in this platform To this end, We identify two distinguishing features between trivial and topological states. Guided by our analytical zero-energy criteria, we find notably that topologically trivial zero-energy states exhibit a zero-bias plateau —rather than a peak— when the magnetic field is swept. This effect is caused by a negative direct-transmission contribution that suppresses the Andreev-reflection peak while maintaining a finite total conductance, stemming from spin-interference effects due to Rashba spin–orbit coupling. 
To our knowledge, this behavior is unique to the topological trivial phase and it has not been observed in the non-trivial phase. Secondly, we find that zero-energy states in both trivial and nontrivial phases exhibit some robustness against disorder. While nontrivial states retreat further into the topological phase, disorder actually expands the range of zero-energy states in the trivial phase.

\section{Acknowledgments}

The authors thank the Deutsche Forschungsgemeinschaft (Project B04, SFB 1277), the Elite Netzwerk Bayern (IGK "Topological Insulators") and the French National Research Agency ANR (project ANR-20-CE30-0028-01) for financial support.
This research was partly funded by the IKUR Strategy under the collaboration agreement
between Ikerbasque Foundation and DIPC on behalf of the Department of Education of the
Basque Government.
H.S. was supported by
Deutsche Forschungsgemeinschaft through CRC 183 (Project No. C03).
\appendix
\section{Zero energy constraints in the lattice model}
\label{appendix: zero energy, TB}
\allowdisplaybreaks
The same approach as in Section \ref{section: zero energy constraints} can be pursued in the discrete model. Here $\cos kd$ is a more convenient variable than $k$. Comparing the coefficients of the two forms of the zero energy polynomial (Eq.~\eqref{eq: zero energy polynomial, coefficients}), where $P(\cos kd)$ is found from Eq.~\eqref{equation: tight-binding dispersion}, we obtain a set of three slightly different constraints, 

\begin{subequations}
 \label{equation: three equations, lattice}
\begin{align}
 O(k^6):\quad & \cosh(qd)\, [\cos (k_f d)+\cos (k_s d)] = \frac{2\tilde{\muTB}}{1+\tilde{\alpha}^2},\\[2mm]
 O(k^4):\quad & \sinh^2(qd)\, [\cos (k_fd) - \cos (k_s d)]^2 = \frac{4\tilde{\alpha}^{2}\tilde{\Delta}^{2}}{(1-\tilde{\alpha}^2)^2}
 \\[2mm]
 O(k^2):\quad &  \sinh^2(qd)+ \cos (k_fd)\cos (k_sd) = \\[2mm]
 & \quad =\frac{\tilde{\muTB}^2 + \tilde{\Delta}^2 - \tilde{V}_Z^2-\tilde{\alpha}^2}{1+\tilde{\alpha}^2},
\end{align}
\end{subequations}
where the tilded quantities are expressed in units of $2t$, $\tilde{x} := x/(2t)$. The different structure of Eqs. \eqref{equation: three equations, lattice} in $\tilde{\alpha}$ compared with the one of $\alpha$ in Eqs. \eqref{equation: three equations, continuum} originates from the changed algebraic relationship between the spin-orbit and hopping contributions: instead of $k, k^2$ in the continuum model, we have $\sin kd, \sqrt{1 - \sin^2 kd}$ (cf. Eqs. \eqref{equation: four bands of the entire hamiltonian}, \eqref{equation: tight-binding dispersion}). Similar to Eqs. \eqref{equation: three equations, continuum}, the conditions in Eqs. \eqref{equation: three equations, lattice} are exact. Keeping now only the terms up to $\sinh^2qd$, we arrive at a constraint on $\Zeeman$:
\begin{align}
 \label{equation: lattice vz constraint, first}
\Zeemansquared& =  \frac{2C_z\left[ (2tz-\muTB)^2 + \alpha^2(z^2-1) + \Delta^2\right]}{2 C_z - (\alpha\Delta)^2} \notag \\
 & - (\alpha\Delta^2) \frac{\left[ (2tz-\muTB)^2 + 8t^2 - 4tz\muTB + \alpha^2(z^2+1) + \Delta^2\right]}{2 C_z - (\alpha\Delta)^2},
\end{align}
where $z:=\cos k_fd$ and $C_z:=[2t(2tz - \muTB) + \alpha^2 z]^2$. When we follow the approximation of Ref.~\onlinecite{DasSarma2012} and drop the terms proportional to $(\alpha\Delta)^2$, the constraint takes a simple form 
\begin{align}
\label{equation: lattice vz constraint, second}
\Zeemansquared = (2t\cos k_fd - \muTB)^2 + \alpha^2(\cos^2k_fd - 1) + \Delta^2,
\end{align}
which upon expansion in $k_f \approx 0$ yields the approximate expression from Ref.~\onlinecite{DasSarma2012} cited in the caption of Fig.~\ref{fig: zero energy lines}.

\section{Energy eigenvalues for limiting cases and open boundary conditions}
\label{section: Eigenvalues of the tight binding Hamiltonian in limiting cases}
We present the results for the lattice Hamiltonian in Eq. \eqref{equation: tight binding Hamiltonian} at $\Delta=0$ in limiting cases considering both the finite length and open boundary conditions. We also shortly comment on the relation to the quantization rule in Eq. \eqref{equation: quantization rule for zero delta}.

\paragraph{$\alpha=0$}: The model in Eq. \eqref{equation: tight binding Hamiltonian} decouples into two separate linear chains for each spin degree of freedom. Thus, the eigenvalues are simply
\begin{align}
	E = -\muTB \pm \Zeeman\,+\,2t\,\cos\left(\frac{n\pi}{N+1}\right),\quad  n=1,\ldots,N,
\end{align}
where "+" ("-") is associated to $\sigma = \uparrow$ ($\sigma = \downarrow$). In the current equation, one finds only two wavevectors associated to the same energy. Thus, we have $k_{1}=k_2$ and $k_1 d = n\pi/(N+1)$ for $n=1,\ldots,N$.

\paragraph{$t=0$}: We find
\begin{align}
	E = -\muTB\pm \sqrt{\Zeemansquared+\alpha^2 \sin^2(k_1 d)}
\end{align}
with $k_1 d +\pi/2 = n\pi/(N+1)$ and $n=1,\ldots,N$. The two signs represent $\sigma = \uparrow,\,\downarrow$ similar to the prior case. The relation between $k_{1,2}$ imposed by Eq. \eqref{equation: tight-binding dispersion} is $k_2 = k_1 -\pi/d$.
\paragraph{$\Zeeman=0$}: The energy eigenvalues are
\begin{align}
	E = -\muTB+2\tilde{t}\,\cos\left(k d\right)
\end{align}
with $\tilde{t} = \sqrt{t^2 + \alpha^2/4}$ and $k d = n\pi/(N+1)$, $n=1,\ldots,N$. Notice that each eigenvalue is twice degenerate.

\section{Momentum quantization in a Dirac system}
\label{appendix: Dirac quantization}
As is already known from the studies of the SSH chain, graphene ribbons and nanotubes, and of the Kitaev chain, in 1D or quasi-1D systems with two (generalized) sites $A$ and $B$ per unit cell the boundary conditions for a system with integer number of unit cells can be set only for one component on one end. For a generic Dirac Hamiltonian
\begin{equation}
H(\mathbf{k}) = \mathbf{k}\cdot\mathbf{\sigma} = \left( \begin{array}{cc}
k_z & k_x-ik_y \\[2mm] k_x+ik_y & -k_z
\end{array}\right)
\end{equation}
the boundary conditions can be set to $\psi_A(0)=0$ and $\psi_B(L) = 0$ (or vice versa, depending on the particular termination). One choice of the eigenvectors diagonalizing this Hamiltonian is 
\begin{eqnarray}
  u_+ = \frac{1}{\sqrt{2\varepsilon(\varepsilon - k_z)}}\left( \begin{array}{c}
  k_x - ik_y \\[2mm] \varepsilon - k_z
  \end{array}\right) , \nonumber \\[2mm] 
   u_- = \frac{1}{\sqrt{2\varepsilon(\varepsilon - k_z)}}\left( \begin{array}{c}
  \varepsilon - k_z \\[2mm] - k_x - ik_y
  \end{array}\right),
\end{eqnarray}
where $\varepsilon = \sqrt{k_x^2 + k_y^2}$ and the energies are $E_\pm(\mathbf{k}) = \pm k$. 
In 1D only one of the $\mathbf{k}$ components remains, and the other two can encode the dependence of the Hamiltonian on external parameters. Let us derive the momentum quantization condition for eigenstates with $E>0$ in a specific case where one of the external parameters is set to 0, so that the Hamiltonian contains only the momentum $k$ and the generalized mass $m$. This yields several possible Hamiltonians, with $(k,m)\in\{ k_x,k_y,k_z\}$. \\
$(i)$ $k_y = k$, $k_x = m$:
\begin{equation}
H(k) = m\sigma_x + k\sigma_y,\quad
u_+ = \frac{1}{\sqrt{2} \varepsilon}\left( \begin{array}{c}
m - ik \\[2mm] |m|
    \end{array}\right).
\end{equation}
The linear superposition of the two Bloch states belonging to $k$ and $-k$ is
\begin{equation}
\psi_+(y) = a_1 u_+(k) e^{ik y} + a_2 u_+(-k) e^{-iky}.    
\end{equation}
Imposing the boundary conditions we find the momentum quantization
\begin{equation}
m = k \cot (kL).    
\end{equation}
This equation has an infinite set of solutions which are neither integer nor half-integer in $\pi/L$ . \\
$(ii)$ $k_y = k$, $k_z = m$:
\begin{equation}
H(k) = k\sigma_y + m\sigma_z,\quad
u_+ = \frac{1}{\sqrt{2|k|(|k|-m)}}\left( \begin{array}{c}
- ik \\[2mm] |k| - m
    \end{array}\right).
\end{equation}
Imposing the boundary conditions on the superposition of Bloch states we find the momentum quantization
\begin{equation}
\cos (kL) = 0.    
\end{equation}
In this case the quantization is $k_n = (n+1/2)\pi/L$.\\
$(iii)$ $k_z = k$, $k_x = m$:
\begin{equation}
H(k) = m\sigma_x + k\sigma_z,\quad
u_+ = \frac{1}{\sqrt{2} \varepsilon}\left( \begin{array}{c}
m - ik \\[2mm] |m|
    \end{array}\right).
\end{equation}
If we imposed the quantization condition in the same way as for the two cases above, we would find
\begin{equation}
  i |m| = k \cot(kL),  
\end{equation}
which clearly cannot be fulfilled for real values of $k$. We find the unphysical values of $k$ because the Hamiltonian itself is not physical -- the diagonal term linear in $k$ indicates that hopping connects the same sites in each unit cell. If both $A$ and $B$ sites support both a forward and backward hopping, the boundary conditions must constrain both components at both ends. If we do that, we recover the usual integer quantization condition $\sin(kL)=0$.

\section{Dirac quantization close to the phase boundary}
\label{appendix: close to TPT}
\begin{figure}[h!]
 \includegraphics[width=\columnwidth]{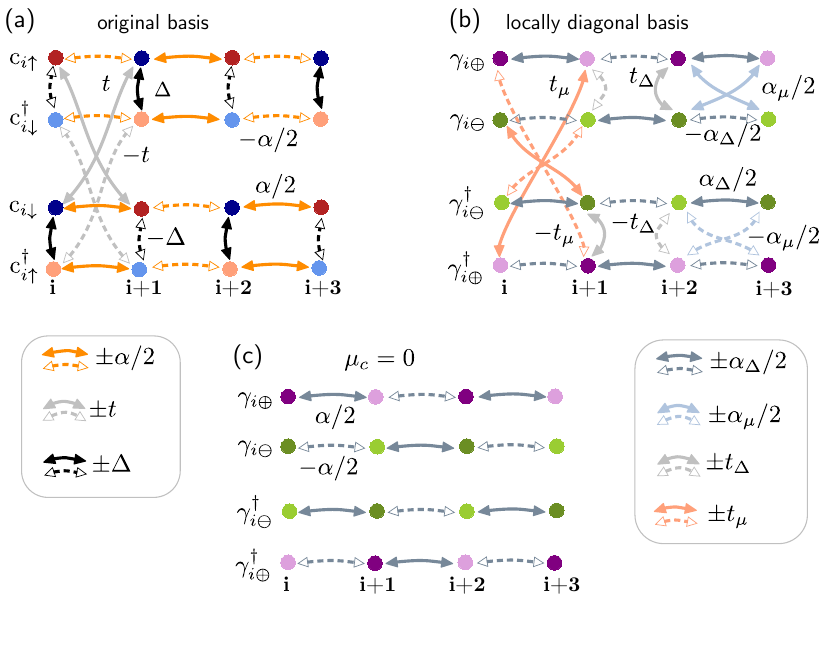}
 \vspace{-10mm}
 \caption{\label{fig:bases}
 Schematics of the non-diagonal Hamiltonian elements in (a) the original and (b) the locally diagonal basis. The basis at site $i+1$ is rearranged to emphasize the decomposition of the system into spin-orbit coupled sub-chains. The last diagram shows the uncoupled subchains for $\muc=0$.
 }
\end{figure}

In order to check whether the quantization close to the phase boundary can indeed be given by $k_n=(n+1/2)\pi/L$, we must check (i) that the dispersion $E(k,\mu)$ or $E(k,\Zeeman)$ with the other parameter fixed is indeed linear, and (ii) that the nanowire Hamiltonian can be decomposed into four chains with two effective sublattices each, and on each end only one sublattice is constrained.~\cite{CastroNeto2009,Marganska2011,Leumer2020}.\\ 
The complexity of the Hamiltonian \eqref{equation: tight binding Hamiltonian} is reduced if we can neglect some of the parameters. In the following we will focus on the parameter region close to the TPT, where the bulk dispersion of the Hamiltonian near the $\Gamma$ point is linear and the spectrum is rich in low energy excitations. \\
If we choose as the Nambu basis in the real space $\boxplus_i(c_{i\uparrow},c_{i\downarrow}^\dag, c_{i\downarrow},c_{i\uparrow}^\dag)^T$, where $\boxplus_i$ denotes here a direct sum over $i$, then
the Bogoliubov-de Gennes form of the Hamiltonian \eqref{equation: tight binding Hamiltonian} can be written as
\begin{equation}
H_{\text{BdG}} = \left( \begin{array}{cccc}
                 D & T & & \\
                 T^\dag & D & T & \\
                  & T^\dag & D & \\
                  & & & \ddots \\
                \end{array}
\right), 
\end{equation}
where
\begin{equation}
D = \left(\begin{array}{cccc}
           \Zeeman - \muTB & -\Delta & & \\
           -\Delta & \Zeeman + \muTB & & \\
           & & -\Zeeman - \muTB & \Delta \\
           & & \Delta & -\Zeeman + \muTB
          \end{array}
\right), 
\end{equation}
\begin{equation}
T = \left( \begin{array}{cccc}
            t & & -\alpha/2 & \\
            & -t & & -\alpha/2 \\
            \alpha/2 & & t & \\
            & \alpha/2 & & -t \\
           \end{array}
\right) .
\end{equation}
The structure of the basis and hoppings is represented schematically in Fig.~\ref{fig:bases}(a). The four operators at each site $i$ are highly interconnected, both on-site and between nearest neighbors. 

In the next step we shall locally decouple the four on-site operators by transforming the basis to one in which the on-site block of the Hamiltonian is diagonal. The transformation is given by
\begin{equation}
\tilde{H}_{\text{BdG}} = U^\dag H_{\text{BdG}} U, \quad\textnormal{where}\quad U = \boxplus_i\left( \begin{array}{cc}
            U_{\oplus i} & \\ & U_{\ominus i} \\
           \end{array}
\right), 
\end{equation}
with
\begin{equation}
U_\oplus = \left( 
\begin{array}{cc}
a_+ & a_- \\
-b_+ & b_- \\
\end{array}
\right),\quad
U_\ominus = \left( 
\begin{array}{cc}
b_- & -b_+ \\
a_- & a_+ \\
\end{array}
\right),
\end{equation}
where $a_\pm := \Delta/\sqrt{2\chi (\chi \pm \muTB)}$, $b_\pm := \sqrt{(\chi \pm\muTB)/(2\chi)}$, and $\chi := \sqrt{\muTB^2 + \Delta^2}$. After the transformation the basis is $\boxplus_i(\gamma_{i\oplus},\gamma_{i\ominus},\gamma_{i\ominus}^\dag,\gamma_{i\oplus}^\dag)^T$, and the Hamiltonian blocks become
\begin{equation}
\tilde{D} = \left( 
\begin{array}{cccc}
\Zeeman + \chi & & & \\
& \Zeeman - \chi & & \\
& & -\Zeeman + \chi & \\
& & & -\Zeeman - \chi
\end{array}
\right), 
\end{equation}
\begin{equation}
 \tilde{T} = \left( \begin{array}{cc  cc}
                     -t_\mu & t_\Delta & \alpha_\mu/2 & \alpha_\Delta/2 \\
                     t_\Delta & t_\mu & -\alpha_\Delta/2 & \alpha_\mu/2 \\[2mm]
                     -\alpha_\mu/2 & \alpha_\Delta/2 & -t_\mu & -t_\Delta \\
                     -\alpha_\Delta/2 & -\alpha_\mu/2 & -t_\Delta & t_\mu
                    \end{array}
\right),
\end{equation}
with
\begin{equation}
X_\mu = X \frac{\mu}{\sqrt{\mu^2+\Delta^2}}, \quad X_\Delta =  X \frac{\Delta}{\sqrt{\mu^2+\Delta^2}},
\end{equation}
where $X\in\{t,\alpha\}$. While we decoupled the four $\gamma$ onsite, we paid for it by complicating the structure of coupling between the sites, illustrated in Fig.~\ref{fig:bases}(b). \\
So far the calculation was general, but now we focus on the parameter space close to the topological phase transition. We set $\muTB=0$ and $\Zeeman = \Delta + v_z$ with small $v_z$ and, because $k$ is small, we neglect the $t$ terms. We find
\begin{align}
    \Tilde{D} & = \left(\begin{array}{c cc c}
    2\Delta + v_z & & & \\
    & v_z & & \\
    & & -v_z & \\
    & & & -2\Delta-v_z
    \end{array}\right),\\[2mm]
    \Tilde{T} & = \left(\begin{array}{c cc c}
     &  & & \alpha/2 \\
     & & -\alpha/2 & \\
    & \alpha/2 & &  \\
    -\alpha/2 & &  & \\
    \end{array}\right).
\end{align}
The spectrum splits into the four subchains illustrated in Fig.~\ref{fig:bases}(c). 
After the Fourier transform and expansion around $k\approx 0$ the Hamiltonian becomes
\begin{equation*}
H(k) = H_\oplus(k) + H_\ominus(k)  ,  
\end{equation*}
where
\begin{eqnarray*}
H_\oplus(k) & \simeq & (2\Delta+v_z)\,\sigma_z - i\,\alpha k\sigma_y,\\[2mm]
H_\ominus(k) & \simeq & v_z\,\sigma_z + i\alpha k \sigma_y.
\end{eqnarray*}
From the sketch in Fig.~\ref{fig:bases}(c) we see that for the four decoupled chains only one of the two components, $\gamma_{i\ominus}$ or $\gamma_{i\ominus}^\dag$, vanishes at each end. The proximitized Rashba nanowire close to the phase transition at $\mu\approx 0$ corresponds to case $(ii)$ from the previous section, with $m=v_z$ and the quantization condition $\cos(kL) = 0$, which indeed results in $k_n = \pi(n+1/2)/L$.
\begin{figure*}[htb]
 \includegraphics[width=\textwidth]{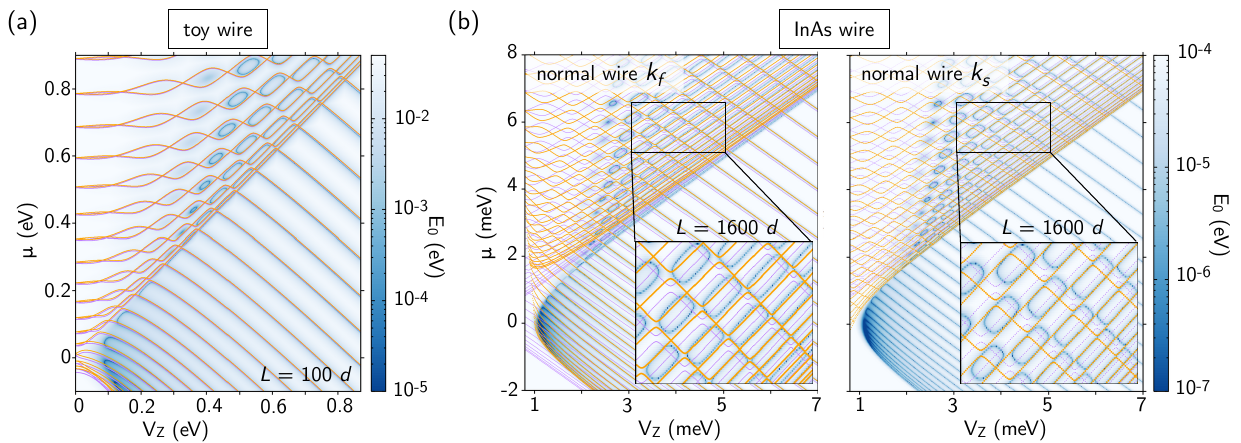}
 \vspace{-5mm}
 \caption{\label{fig: zero lines - normal spectrum}
 Energy of the lowest excitation (color background), spectrum of the normal wire (purple lines), and approximate analytical zero energy lines with either (a),(b) $k_f$ or (c)$k_s$ taken from the normal wire (orange lines). For the spectrum the vertical axis is $\muc=E^{\Delta=0}$.
 }
\end{figure*}

\section{Relation between the normal wire spectrum and the zero energy lines}

One of the natural questions arising from reading Sec.~\ref{section: finite} is that of the relationship between the spectrum of the normal wire and the presence of zero energy modes when the wire is proximitized. The addition of the superconducting pairing $\Delta$ modifies the boundary conditions, entangling the electron and hole degrees of freedom. The relation of $\Delta$ to the other energy scales -- $\muc,\Zeeman$ and the size quantization $\delta \varepsilon:=\hbar^2(\pi/L)^2/(2m)$ is a crucial factor here, as can be seen in Fig.~\ref{fig: zero lines - normal spectrum}. In a short wire, where the size quantization of the normal wire is the dominant energy scale, the non-superconducting energy levels (shown as purple lines) coincide almost perfectly with the zero energy lines in the whole $(\Zeeman,\muc)$ plane; they fail only to reproduce the horizontal spacing between the zero energy rings in the trivial phase. The wavevectors of the superconducting wire are determined mostly by those of the normal wire.\\
In the long wire $\delta\varepsilon < \Delta$ and at low $\muc,\Zeeman$ the boundary condition  is strongly modified by the superconducting pairing. The spectrum of the normal wire begins to coincide with the zero energy features only at high $\Zeeman$, where the Zeeman effect dominates over the other energy scales.\\
The orange lines in Fig.~\ref{fig: zero lines - normal spectrum}(b),(c) show the complementary nature of $k_f$ and $k_s$ in the spectrum of the normal wire. If $k_f$ is taken from the normal wire and $k_s$ adjusted to it, the condition in Eq.~\eqref{equation: lattice vz constraint, first} follows correctly the zero enery lines in the topological phase and the downsloping boundaries of zero energy rings. When in the trivial phase we choose instead $k_s$ and adjust $k_f$, it is the upsloping boundaries which are well reproduced. 

\bibliographystyle{apsrev}
\bibliography{references} 
\end{document}